# HEK-Omics: The promise of omics to optimize HEK293 for recombinant adeno-associated virus (rAAV) gene therapy manufacturing

Sai Guna Ranjan Gurazada[1], Hannah M. Kennedy[2], Richard D. Braatz[3], Steven J. Mehrman[4], Shawn W. Polson[1*], Irene T. Rombel[2*]

[1]Center for Bioinformatics and Computational Biology, Department of Computer and Information Sciences, University of Delaware, Newark, DE, United States

[2]BioCurie, Inc., Wilmington, DE, United States

[3]Massachusetts Institute of Technology, Cambridge, MA, United States

[4]Janssen Research & Development, Janssen Pharmaceuticals, Johnson & Johnston, Malvern, PA, United States

*To whom correspondence should be addressed: Irene T. Rombel (irene@biocurie.com) and Shawn W. Polson (polson@udel.edu)

*Irene Rombel will handle correspondence for the article at all stages of the refereeing and publication process, and post-publication* (irene@biocurie.com).

## Abstract

Gene therapy is poised to transition from niche to mainstream medicine, with recombinant adeno-associated virus (rAAV) as the vector of choice. However, this requires robust, scalable, industrialized production to meet demand and provide affordable patient access, which has thus far failed to materialize. Closing the chasm between demand and supply requires innovation in biomanufacturing to achieve the essential step change in rAAV product yield and quality. Omics provides a rich source of mechanistic knowledge that can be applied to HEK293, the prevailing cell line for rAAV production. In this review, the findings from a growing number of disparate studies that apply genomics, epigenomics, transcriptomics, proteomics, and metabolomics to HEK293 bioproduction are explored. Learnings from CHO-Omics, application of omics approaches to improve CHO bioproduction, provide context for the potential of "HEK-Omics" as a multi-omics-informed approach providing actionable mechanistic insights for improved transient and stable production of rAAV and other recombinant products in HEK293.

## 1. Introduction

Gene therapies have the unique ability to prevent, treat, or cure many of the ~7000 rare hereditary diseases that afflict over 30 million Americans (Mendell et al., 2021; Wang et al., 2019), and 400 million people worldwide ("Rare diseases," 2022). Recombinant adeno-associated virus (rAAV) is a potent tool for the delivery of genes into mammalian cells for therapeutic benefit (Chancellor et al., 2023; Kuzmin et al., 2021; Wang et al., 2024), and has emerged as the most common vector for *in vivo* gene therapy, as evidenced by a growing list of FDA-approved products (Center for Biologics Evaluation and Research, 2023) and more than 250 candidates currently in the clinic (Burdett and Nuseibeh, 2023; "Regenerative medicine: Disrupting the status quo," 2022).

While the clinical and commercial opportunity is large, rAAVs are challenging to manufacture because of their high complexity, resulting in exorbitant costs, long lead times, quality issues, and vector shortages that contribute to clinical trial and commercial delays (Dobrowsky et al., 2021; Escandell et al., 2022; "High-dose AAV gene therapy deaths," 2020; Mullard, 2021; Sha et al., 2021; Srivastava et al., 2021). There is an acute need for scalable, robust, and cost-effective processes compatible with industrial-scale production of gene therapies, reminiscent of a comparable need faced by the biopharmaceutical industry in the early days of therapeutic antibody production (Li et al., 2010).

Human embryonic kidney (HEK) 293 cell lines are versatile systems used for the production of recombinant proteins, viral and virus-like particles. With advantages that include a high transfection rate, fast growth rate, and human-like post-translational modification, HEK293 is the most prevalent host cell for the manufacture of rAAV vectors (Dobrowsky et al., 2021; Naso et al., 2017; Srivastava et al., 2021). The original parental immortalized cell line was created in 1973 by the integration of a 4 kbp adenoviral 5 (Ad5) genome fragment encoding E1A and E1B genes into cells derived from the kidney of an aborted human embryo (Malm et al., 2020). In addition to enabling the continuous culturing of HEK293 (Berk, 2005), E1A and E1B gene products are necessary helper proteins for rAAV production (Qiao et al., 2002). Since its creation, many derivative cell lines with a range of advantages over the parental HEK293 have been developed (Zhang et al., 2024).

HEK293 cell lines are second only to Chinese hamster ovary (CHO) cells for their use in biopharmaceutical production, and are the prevailing system for small-scale protein and viral vector propagation (Lin et al., 2014; Tan et al., 2021). As with the development of CHO to meet the demands of therapeutic antibody production, improving HEK293 host cell productivity through innovative engineering strategies will be critical to meet the growing clinical and commercial need for safe, potent, and cost-effective rAAV products (Escandell et al., 2022; Sha et al., 2021). Although the upper theoretical limit for rAAV production in HEK293 is not known, comparison with monoclonal antibody production in CHO cells implies a manufacturing productivity gap in the order of 1000 fold (Wright, 2023).

Beyond process enhancements to improve product quality, efficiency, and scale – and stable producer cell lines to simplify the ubiquitous transient triple transfection step (Dobrowsky et al., 2021; Naso et al., 2017; Srivastava et al., 2021) – mechanistic models have been developed and applied to significantly enhance processes within this step, further improving product quality and yield (Canova et al., 2023; Destro et al., 2023; Nguyen et al., 2021). The host cells are a critical raw material that governs overall productivity, and much attention is needed to ensure optimal performance.

Systems biology provides an approach to identify and alleviate bottlenecks, specifically those within the cellular machinery involved in rAAV production. Individual omics technologies such as genomics, epigenomics, transcriptomics, proteomics, and metabolomics can provide valuable insight into the optimal environment for the cells and cellular mechanisms involved in rAAV production by HEK293 cells. In concert, they provide a synergistic multi-omic systems biology approach that could enable significant gains through data-driven engineering or modeling. The CHO field provides a precedent to build on: CHO-Omics (Kildegaard et al., 2013; Dahodwala and Sharfstein, 2017; Lin et al., 2020; Sorourian et al., 2024; Stolfa et al., 2018) which enabled the mature development of this cell system to meet the needs of many commercial products. We review the field's current state in respect to omics applications and identify gaps and future prospects which support "HEK-Omics" as a multi-omics-informed approach to HEK293 pharmacological modulation or engineering for improved production of rAAV and other recombinant products.

## 2. Genomics

Genomics provides unique insight into the genetic makeup of prokaryotes, eukaryotes, and viruses (Houldcroft et al., 2017; Konstantinidis and Tiedje, 2005; Sharma et al., 2018). Cellular activity is constrained by contents and organization of the genome, representing the baseline for what can be accomplished within a cell to make it a suitable host for use in biotherapeutics (Wuest et al., 2012). Historically, measures of a cell's genome have been difficult. Over the past two decades, high-throughput DNA sequencing has evolved through multiple generations of technological advancements driving commensurate decreases in sequencing cost which have facilitated broad access to genomics approaches including *de novo* genome sequencing and genome resequencing, and led to development of approaches enabling exploration of the transcriptome and epigenome (Giani et al., 2020; Satam et al., 2023). Consequently, application of genomics has expanded from basic research to a multiplicity of industrial and clinical applications, from disease diagnosis to the design and implementation of precision medicines (Denny and Collins, 2021; Green et al., 2020; Lähnemann et al., 2020). In this section, we review how genomics has been applied to HEK293 cells in the context of biotherapeutic production.

*2.1. De Novo Genome Assemblies Provide a Foundation for HEK-Omics*

DNA sequencing has been used to establish reference *de novo* genome assemblies that are foundational to unraveling the molecular characteristics of any organism. For HEK293 cells, the sequencing of the human genome (Venter et al., 2001) and subsequent improvements to this reference assembly (Singh et al., 2022) have paved the way for HEK-Omics research. The current human reference genome version, GRCh38 (aka hg38), established by the Genome Research Consortium (GRC) in 2017, contains 2.92 Gbp of sequence composed of 24 chromosomes including chromosomes X and Y, and contains 60,090 identified genes including 19,890 protein-encoding genes (Nurk et al., 2022; Schneider et al., 2017). In 2022, the Telomere-to-Telomere (T2T) consortium published the first "completely assembled", gapless human reference genome version T2T-CHM13, for all 22 human autosomes and chromosome X, comprising 3.05 Gbp of nuclear DNA, plus a 16.5 kbp mitochondrial genome. The draft T2T-CHM13 annotation describes 63,494 genes, of which 19,969 genes are predicted to encode proteins (Nurk et al., 2022). While either version can be used as a reference genome for HEK293 sequence analysis, the GRC reference series is more widely adopted (Chung et al., 2023; Malm et al., 2022, 2020; Saghaleyni et al., 2022), primarily due to its well-established suite of ancillary databases and bioinformatics resources (Frankish et al., 2021; Lee et al., 2020; Yates et al., 2020).

*2.2. Genome Resequencing to Characterize and Adapt HEK293 for Production*

Genome resequencing helps elucidate the evolution of an organism or cell line by characterizing the genetic changes accumulated over time due to genetic drift or adverse selective pressure. The genetic changes are typically characterized in the form of single nucleotide variants (SNVs) and polymorphisms (SNPs), insertion/deletion variants (InDels), structural variations (SVs), copy number variations (CNVs), and chromosomal recombinations.

Using whole genome resequencing, five widely used rAAV-producing HEK293 cell lines (293T, 293S, 293SG, 293SGGD, 293FTM) have been characterized, by comparing genetic variations of the parental HEK293 cell line relative to the human reference genome (Lin et al., 2014). CNV analysis identified key genomic regions driving HEK293 cell adaptation. The high copy number of the adenovirus insertion site, potentially coupled with low copy number of the fumarate hydratase (FH) gene locus, appears to be a major factor in the transformed phenotype. While these CNVs help explain diverse phenotypes across HEK293 subclones, further experimentation is needed to confirm their roles in suspension growth adaptation, cellular transformation, and metabolism.

Many production cell lines, including CHO and HEK293, were initially developed as adherent cell lines with surface attachment requirements. However, adherent cell culture systems are a challenge for scale-up to meet the high demand required for commercial supply, motivating the adaptation of adherent cell lines to suspension cell lines (Chu and Robinson, 2001). As part of a larger multi-omics study, genomics has also been used to analyze a set of HEK293 adherent (293, 293E, 293T) and suspension (293H, 293F, freestyle 293F) cell culture lines used in recombinant protein production (Malm et al., 2020) (see Sections 4, 6, and 7 for further findings on this study). Hierarchical clustering of the genomic data revealed (1) a significant divergence of all the derivative cell lines from the parental HEK293 cell line, and (2) that the adherent cell lines clustered together a single group that was distinct from the group of suspension cell lines. Pairwise comparison of SNVs and InDels between the HEK293 progeny cell lines and the parental line revealed that two adherent cell lines, HEK293E and 293T, had the highest number of high- and moderate-impact SNVs, the type of variants most likely to disrupt protein function. Five genes in particular (CYFIP2, C9orf43, PPP2R4, SGCD, and CTB-47B11.3) had acquired high-impact SNVs across all the progeny cell lines, and could be potential candidates for targeted cell line engineering. CNV analysis identified similar patterns of copy number gain or loss across the progeny cell lines, with an ~15 Mb region on chromosome 13 consistently showing copy number gain in all progeny cell lines. Understanding these fundamental genetic traits characteristic of the producing cell lines, while distinguishing the adherent lines from the suspension cell lines, is a critical first step in helping explain the phenotypic and physiological differences between the commonly used rAAV-producing HEK293 cell lines. The aforementioned studies highlighted a number of opportunities to enhance suspended HEK293 cells. The suspension cell group displayed behavior directing enhancement of adherent properties (cellular adhesion), presumably inherited from progeny cell lines. Re-directing this effort can lead to benefits of overall efficiency and promote better cellular suspension properties. For example, it was found that cholesterol content and distribution strongly relate to secretory mechanism optimization, facilitating Golgi transport (Malm et al., 2020). By experimentally establishing a clear association of genomic markers to desirable phenotypic traits, this could inform the engineering of improved transient and stable producer cell lines in HEK293.

Not only does genome resequencing apply across cell line derivatives, but also within the same cell line to determine the effects of cell procedures on the genome. There has always been a concern whether standard cell cultivation procedures such as cell banking and cultivation introduce genomic instability, which can, in turn, cause phenotypic instability (Borsi et al., 2023; Wurm and Wurm, 2017). CHO, for example, has a propensity for genomic rearrangement that, in combination with the genomic and metabolic demand of high-producing cells, manifests as a decline in productivity and product quality (Dahodwala and Lee, 2019). HEK293, in contrast, does not appear to share this degree of genomic instability. Upon comparing the HEK293T cell line (genome with highest number of SVs) to the genomic samples of the same cell line after multiple rounds of cultivation, banking, freezing, and thawing, it was established that the genome of these cells is stable throughout standard cell culturing conditions, as the samples were very similar in terms of gene copy numbers, point mutations, and SVs (Lin et al., 2014). The HEK293S line was an exception, as dramatic copy number changes were detected upon thawing. Nevertheless, genomic instability is observed in HEK293 cell lines under selective conditions as evidenced by larger chromosomal rearrangements (for example, an 800 kb deletion was detected by Lin et al. (2014) in their analysis of SVs) rather than point mutations in the ethyl methanesulfonate-mutagenized line, HEK293SG. Such efforts to examine the genomic characteristics of HEK293 cell lines and their evolution under selective pressure are an important step towards explaining performance variation in producing cell lines, and providing valuable insights for rational engineering of high-producing cell lines that are genetically stable.

While biomedical study has extensively leveraged genomics in humans and cell lines modeling specific human health conditions to provide increasing understanding of genetic variability on biological processes *in vivo* – this represents only a starting point for understanding and optimizing cell lines leveraged in a production setting. Targeted studies toward understanding the genotypic basis of favorable production phenotypes provide crucial data to help us understand the basis of existing limitations and will inform future directed evolution and bioengineering efforts to address the existing productivity gap in HEK293 production systems.

## 3. Epigenomics

Epigenomics is the study of epigenetic changes in the form of chemical modifications to a cell's genomic DNA and/or chromatin that have the ability to regulate gene activity. Derived from the latin term "epi", which means "on top of", epigenomics is genomics that goes beyond the primary nucleotide sequence. Epigenetic changes can be inherited from parental cells, or acquired as a result of direct environmental exposure to nutritional, chemical, and physical factors, and can influence phenotypic variability independent of genotypic variability (Cavalli and Heard, 2019; Petronis, 2010). In contrast to genomics, which is typically defined by a fixed primary DNA sequence, epigenomics is a reversible process that affects gene expression without altering the underlying base pair sequence. The best characterized epigenetic changes include DNA base modifications (e.g., methylation of cytosine residues primarily at CpG sites) and chromatin remodeling due to modification (e.g., acetylation) of histone proteins (Allis and Jenuwein, 2016; Jeltsch et al., 2018; Zhang et al., 2020). In this section, we review evidence of epigenetic activity in HEK293 cells and its potential effect on the production of recombinant proteins. Harnessing and controlling this behavior is key to building scalable and high-producing cells to deliver optimal processes to support clinical trials and commercial supply.

### *3.1. Epigenomics Analysis Techniques*

Similar to other omics approaches, several high-throughput molecular techniques have been developed for epigenomic analysis (Rivera and Ren, 2013; Turunen et al., 2018). The most commonly used techniques to characterize each epigenetic modification are genome-wide DNA methylation assays such as bisulfite sequencing (BS-seq); histone modification assays such as chromatin immunoprecipitation followed by sequencing (ChIP-seq); chromatin accessibility assays such as micrococcal nuclease digestion with deep sequencing (MNase-seq), DNase I hyper-sensitive sites sequencing (DNase-seq), formaldehyde-assisted identification of regulatory elements followed by sequencing (FAIRE-seq) or assay for transposase-accessible chromatin using sequencing (ATAC-seq); and finally chromatin interaction assays such as Hi-C, which is the genome-wide version of the chromatin conformation capture (3C) method (Mehrmohamadi et al., 2021). More recently, epigenomic studies at the level of individual cells, rather than bulk populations, have also been possible through respective single-cell (sc) epigenomic assays such as scBS-seq, scChIP-seq, scATAC-seq, and scHi-C (Mehrmohamadi et al., 2021; Wang and Chang, 2018).

### *3.2. DNA Methylation Reveals Potential Epigenetic Targets*

In a study to assess the impact of genome-wide epigenetic modifications in HEK293 cells, DNA methylation was introduced at several thousand CpG islands in a single experiment (Broche et al., 2021). These epigenetic modifications were shown globally to strongly correlate with a decrease in gene expression, suggesting direct impact. Transient reduction in methylation was also observed globally for H3K4me3 and H3K27ac, which are important histone tags and considered as transcription activation epigenetic biomarkers (Igolkina et al., 2019). Notably, H3K4me3 is associated with a plethora of histone modifications, and genes marked with the associated H3K4me3 domain are involved in cell identity and essential cell functions (Beacon et al., 2021). Likewise, H3K27ac is an important marker that can distinguish between active and inactive/poised enhancer elements (Creyghton et al., 2010).

As a corollary, hypomethylation of host cell DNA by treatment with 5-Aza-2’ deoxycytodine, a hypomethylation agent, was shown to have a positive influence on wild-type AAV (wtAAV) integration and rAAV transgene expression (Chanda et al., 2017). Although the study was performed in HeLa cells with the goal of assessing the potential impact of in vivo epigenetic events in the human host (i.e., the patient), this study also sheds light on the impact of cellular methylation in human-derived cell lines including HEK293. It is evident from these and other studies that epigenetic targets present a viable and as yet untapped opportunity for HEK293 engineering to improve rAAV and other recombinant protein production.

### *3.3. Production Stability Assessed by Epigenomics*

Phenotypic stability, particularly production stability, is a prerequisite for biopharmaceutical manufacturing. However, unexpected and uncontrolled phenotypic variation remains a major challenge, and is not always explained by genetic variability alone (Moritz et al., 2016, 2015; Osterlehner et al., 2011; Paredes et al., 2013; Patel et al., 2018), implying a potential role of epigenetic variability. Understanding the role of epigenetic variations in a production cell line (and processes therein) can inform engineering strategies to improve stability through epigenetic regulation (Marx et al., 2022). Additionally, epigenetic regulation can also be responsible for the adaptation of mammalian cells to growth or media conditions that are more conducive to large-scale manufacturing, such as transition from adherent to suspension culture or serum-replete to serum-free media (Malm et al., 2020; Mcallister et al., 2002). To investigate these aspects, DNA methylation, histone modification, and genomic data from six related CHO cell lines, subjected to either long-term cultivation (maintained in culture for 3 to 6 months) or different evolutionary pressures (suspension growth, glutamine-free medium), were collected and the genomic and epigenomic modifications were analyzed (Feichtinger et al., 2016). Bisulfite sequencing was used to estimate genome-wide methylation profiles, and ChIP-seq was used to evaluate chromatin states and modifications. Only minor changes in DNA methylation patterns were observed when cells were cultivated under the same conditions over prolonged periods; however, methylation patterns correlated with changes in culture conditions such as a glutamine-free medium or culture methods for inducing adherent to suspension phenotype switch. This finding suggests that epigenetic regulatory mechanisms are involved in helping cells overcome evolutionary pressures, and thus could be used as potential targets for epigenetic editing and cellular engineering to engineer robust and highly productive host cell lines.

### *3.4. Production Stability and Safe Harbor Regions*

When establishing a producing cell line, production instability is often caused by random integration of transgenes into locations on the host cell genome that are vulnerable to genetic and epigenetic instability (Kim et al., 2011; Osterlehner et al., 2011; Wippermann and Noll, 2017; Wurm, 2004). To investigate this phenomenon in CHO cells, the epigenomes and transcriptomes of industrially relevant monoclonal antibody-producing CHO cell line and its parental CHO-K1 host have been extensively characterized to identify safe harbor regions (Hilliard and Lee, 2021). 'Safe harbor region' is a term used to describe genomic loci that are transcriptionally permissive and have enhanced stability. The Hi-C technique for chromatin structure analysis and RNA-seq for system-wide gene expression analysis were used to establish that 10.9% of the CHO genome was marked by transcriptionally permissive chromatin structures with enhanced stability relative to the rest of the genome. During a typical cell line development process, that involves site-specific integration of transgenes and/or regulatory elements, identification of these safe harbor regions dramatically reduces the search space for safe chromosomal locations, and could potentially minimize production instability.

Similar productivity and scale-up challenges prevail in biotherapeutic manufacturing using the HEK293 host cellular system; therefore, utilizing the decades of learnings from CHO platforms, comprehensive epigenomic studies in HEK293 systems can broaden their therapeutic application (Pulix et al., 2021). Epigenomic studies allow for characterization of the HEK293 cell line and its variants to assess the safe harbor regions for stable gene expression. For example, three human genomic safe harbors (AAVS1, CCR5, and human ROSA26 loci) were evaluated with respect to the transgene expression level and stability in HEK293 cells (Shin et al., 2020). Upon evaluation, AAVS1 showed the highest homogeneity and expression of enhanced green fluorescent protein (EGFP). This then led to the creation of a novel cell line using CRISPR/Cas9-mediated integration of the landing pad into AAVS1 locus. The resulting cell line displayed faster generation of recombinant cell lines that produce therapeutic proteins to increase productivity and aid scale-up challenges that prevail in the HEK293 host cellular system.

Epigenetics may be a cause of unexplained outcomes that hamper high, consistent production in HEK293. For example, the reduction of cell specific productivity at increasing cell densities – also known as the cell density efficient (CDE) – might be explained by certain genes being silenced or turned on inconsistently (Lavado-García et al., 2022, 2020b). However, the role of epigenetics in these unexplained phenomena has not been explored to date. Herein lie opportunities to investigate the role of epigenetic regulatory mechanisms, in combination with other omics approaches, which provide the opportunity to unlock the productivity gains for large-scale manufacturing of rAAV and other biotherapeutics.

## 4. Transcriptomics

Transcriptomics is a field in which researchers study the complete set of all RNA transcripts produced by the genome and expressed in a given cell, tissue, or organism (Morozova et al., 2009; Wolf, 2013). Transcriptomic analysis has enabled the study of how gene expression changes in an organism under external or internal stresses, thereby giving insight into the organism's biology and gene regulatory networks at play (Lowe et al., 2017). Researchers that implement genetic engineering to study and evaluate phenotypic control for example, can utilize transcriptomics to evaluate the regulatory networks at play for gene expression at the transcriptional, post-transcriptional, translational and post-translational layers (Eisenhut et al., 2024). In this section, we review how transcriptomics can be implemented to gain insights into host cell limitations, targeting, and virus or protein secretion pathways.

### *4.1. Transcriptomics Analysis Techniques*

Transcriptomics typically uses two technologies: microarrays to quantify the expression of a set of predetermined sequences, and RNA-seq to capture the expression of all genome-wide transcripts (Lowe et al., 2017; Wang et al., 2009). Micro-arrays are a high throughput technique that can quantify the predetermined sequences via their hybridization to an array of complementary probes. There is a vast range of genes covered in the micro-arrays available. On the other hand, RNA-seq delivers valuable information into the complexity of transcriptomes by providing high-resolution and global measurement of transcript abundance and their isoforms through sequencing of reverse transcribed cDNA.

### *4.2. Host Cell Limitations*

Transcriptomics has been instrumental in identifying limitations in host cell systems producing recombinant proteins for therapeutic applications. For example, gene expression dynamics within the secretory pathways of the host cells, transiently expressing CHO and HEK293, were compared during the production of 24 difficult-to-express proteins (Malm et al., 2022). Pathway enrichment analysis of differentially expressed genes in CHO and HEK293 revealed that (1) HEK293 cells showed higher expression for genes involved in protein secretion, (2) protein folding genes were significantly suppressed in CHO cells while being significantly upregulated in HEK293 cells, and (3) both showed significant upregulation of genes involved in translation, as evidenced by a spike in ribosomal expression. In some cases, post-translational modification (PTM) machinery was observed to be key to the productivity of recombinant proteins in the two cell lines. For example, the higher titer in HEK293 for more heavily glycosylated products was associated with increased activities of N- and O-glycosyltransferases in HEK293 compared to CHO. Rumachik and colleagues showed that rAAV capsids have PTMs, including glycosylation, methylation, acetylation, and phosphorylation, that differ between production platforms (Rumachik et al., 2020). While the clinical implications of these PTMs are unknown, understanding these dynamics and their ability to control PTMs will inform the development of more robust processes and cell lines and potentially safer, more potent gene therapy products.

### *4.3. Cellular Targets for Improving Production*

Transcriptomics has been used for characterization of HEK293 derivatives, providing insights into molecular mechanisms that correlate with the varying phenotypic and morphological profiles of different HEK293 cell lines (Lin et al., 2014; Malm et al., 2020; Saghaleyni et al., 2022). As part of a larger multi-omics study, transcriptomics has been used to compare widely used HEK293 adherent cell lines with suspension cell lines (Malm et al., 2020) (see Sections 2 and 6 for more findings on this study). The differential gene expression and functional enrichment analysis between the two groups revealed that genes involved in cellular compartment organization, cell adhesion, cell differentiation, cell morphogenesis, and cell motility, but not in extracellular matrix organization were significantly elevated in suspension cell lines. Gene enrichment analysis further pointed to the cholesterol biosynthesis pathway as significantly enriched among the differentially expressed genes between adherent and suspension cells. Compared to adherent cell lines, five key differentially expressed genes – Lysyl Oxidase (LOX), Inhibitor of DNA Binding 1 (ID1), Zic Family Member 1 (ZIC1), Dehydrogenase/Reductase 3 (DHRS3), and Retinoic Acid Receptor Gamma (RARG) – were consistently downregulated in suspension cell lines. These results are in agreement with public transcription data collected from 47 adherent and 16 suspension human cell lines in the Human Protein Atlas database (http://www.proteinatlas.org). If linked to morphological variation between adherent and suspension cell lines, these genes could be targets for cellular engineering or pharmacological modulation.

Five widely used HEK293 cell derivatives (293-T, 293-S, 293-SG, 293-SGGD, and 293-FTM) have also been compared to each other and the parental HEK293 cell lines (Lin et al., 2014). Transcriptomics revealed differential gene expression consistent with cell cycle activation and proliferation, potentially due to the faster growth rate of the derivative cell lines compared to the parental HEK293 line. Additionally, most of the 136 genes differentially expressed between derivative cell lines and the parental HEK293 cell line were found to be involved in cell adhesion and motility. This is potentially indicative of the culture characteristics of the parental adherent HEK293 cell line, which is known to be challenging to physically dissociate from culture dishes compared to its derivatives. Discovering such molecular mechanisms and key transcriptomic differences between the non-producing parental HEK293 cell line and the derivative lines, could lead to new hypotheses and targets for engineering better production cell lines faster.

In addition to finding new targets, transcriptomics has helped elucidate the underlying molecular mechanisms for validated target(s) that led to productivity improvement in HEK293. For example, when Caspase 8 Associated Protein 2 (CASP8AP2) gene was identified as a potential target for engineering improved expression of recombinant protein in HEK293 from a genome-wide siRNA screen (Xiao et al., 2016), the knockout of CASP8AP2 gene led to a 7-fold improvement in constitutive expression of recombinant protein luciferase and a 2.5-fold increase in transient expression of recombinant protein SEAP in HEK293 (Abaandou et al., 2021). Transcriptomics was used to compare the CASP8AP2-knockout cell lines to the parental lines to reveal that the cell cycle regulation pathways were the most prominent among the top 25 differentially expressed pathways. This further led to the identification of the Cyclin Dependent Kinase Inhibitor 2A (CDKN2A) gene and its significant upregulation, as a key factor for improving recombinant protein expression in HEK293 cells.

### *4.4. Bottlenecks in Virus and Protein Secretion*

To identify bottlenecks in protein secretion for improving recombinant protein production, transcriptomics was used to compare HEK293F stable cell lines expressing either the non-secretory green fluorescent protein (GFP) or the secretory protein erythropoietin (EPO) at varying translational rates (Saghaleyni et al., 2022). EPO producers showed a higher ATP production capacity, indicated by upregulated genes in the oxidative phosphorylation pathway. Additionally, ribosomal gene expression patterns varied based on the type and rate of recombinant protein production. For example, the expression of mitochondrial ribosomal genes was positively correlated to EPO secretion; however, no such evidence was found with GFP production, whereas the expression of cytosolic ribosomal genes was negatively correlated to production rate in both cases. Finally, high EPO-producing clones significantly upregulated the ATF6B gene, potentially enhancing the ER stress response to manage high protein secretion. ATF6B is a transcription factor known to be active under ER stress conditions due to accumulation of unfolded proteins. The gene is normally integrated within the ER membrane but enters the nucleus to activate ER stress response genes under stress conditions (Iurlaro and Muñoz-Pinedo, 2016; Thuerauf et al., 2004). Understanding the effects of cargo proteins on gene expression and protein secretion can guide cargo engineering to enhance protein secretion.

Understanding the cellular response of HEK293 cells to rAAV production, and identifying genes and pathways that influence rAAV production, is pivotal to developing robust manufacturing platforms that provide safe, scalable, and affordable rAAV-based gene therapies. An antiviral response has been shown to affect protein synthesis, cellular metabolism, cell proliferation, and even programmed cell death (Fritsch and Weichhart, 2016), and could be a significant factor in curtailing rAAV production in HEK293 cells. To understand the underlying mechanism, RNA-seq was used to interrogate the transcriptional response of suspension HEK293 cells to rAAV production following transient transfection, under manufacturing-relevant conditions (Chung et al., 2023). Systematic analysis indicated that vector manufacture actively triggers an innate infectious response in the host cells, as evidenced by the upregulation of pathways involving antiviral and inflammatory responses. In an orthogonal study, transcriptomic comparison between rAAV-producing and non-producing HEK293 cultures over time revealed that the antiviral innate immune response is a significant bottleneck in rAAV production, and could be addressed by supplementing the medium with small-molecule inhibitors, or by gene silencing or knockout (Wang et al., 2023). J. Fraser Wright postulates that the production efficiency gap between rAAV and monoclonal antibodies could be, in part, explained by the difference in innate antiviral activity in human-derived HEK293 cells and non-human derived CHO cells (Wright, 2023). Hence developing mechanisms to limit these responses in host cells could potentially close the gap and help improve rAAV production.

### *4.5. Identification of Epigenetic and Cell Cycle Regulators and Druggable Targets*

To gain a better mechanistic understanding of key pathways during rAAV production, a transcriptomic analysis combined with pharmacological perturbation was applied to five HEK293 cell lines with variable capacities for rAAV production (Tworig et al., 2024). An RNA-Seq time-course was used to identify genes that are both differentially expressed between base and high rAAV producers, and modulated during rAAV production. This revealed core cellular functions that are upregulated (e.g., cellular response, signal transduction, cell cycle regulation), or downregulated (e.g., mitochondrial electron transport, glutathione metabolic modulators, proteasomal proteins), reflecting a complex shift in cell state, with widespread modulation of energy production and biosynthesis. With this knowledge of potential druggable pathways and targets, small molecule compounds were selectively tested. Up to 19% of the tested compounds improved rAAV production by 1.5-fold above baseline, which is substantially higher than the typical hit rate of 0.1–1% using an unbiased high-throughput screen. Of the drug classes that significantly enhance rAAV production, those targeting epigenetics, transcriptional activation, and cell cycle modulation were particularly compelling, Notably, two HDAC inhibitors produced titers of more than 4-fold and 9-fold above baseline, underscoring the importance of epigenetics and chromatin remodeling in regulating rAAV expression and productivity in HEK293.

Transcriptomics has enabled a comprehensive view of gene expression dynamics, contributing to our understanding of cellular functions, protein and virus secretion, and potential therapeutic targets, paving the way for actionable insights to enhance HEK293 performance.

## 5. Proteomics

The proteome describes the set of proteins encoded by the genome (Wilkins et al., 1996). Proteomics is the study of the functional proteome that enables identification of expressed proteins at a given time, and sheds light on their function and structure, including spliced protein isoforms, post-translational modifications (PTMs), interaction partners, structural description, and higher order protein complexes (Tyers and Mann, 2003). While genomics provides the blueprint of all possible gene products, proteomics builds on this foundation to provide the functional context for the translated gene products. Proteomics is complementary to genomics and transcriptomics, and together can be used to model cellular behavior at the whole system level, and are key tools in the study and optimization of systems biology. In this section, we review how proteomics can be used for enhancing host cell productivity through expression profiling, PTM discovery, protein activity, and analysis of cell composition and organelle composition.

### *5.1. Techniques for Proteomic Analysis*

The most common techniques for protein identification and characterization in mammalian cells (Farrell et al., 2014) are (1) antibody-based separation methods such as western blotting (Kurien and Scofield, 2015) and enzyme-linked immunosorbent assay (ELISA) (Lequin, 2005), and (2) gel-based separation methods such as two-dimensional gel electrophoresis (2DE) (Dunn, 1987) and two-dimensional difference in-gel electrophoresis (2D-DIGE) (Marouga et al., 2005). These techniques are considered low throughput, only able to identify and characterize tens to hundreds of proteins. Accordingly, for a robust proteomics analysis, a variety of high-throughput techniques based on mass spectrometry (MS) have emerged. These methods allow not only for protein identification and characterization, but also relative and absolute protein quantification (Ankney et al., 2018). Among these techniques are gel-free separation methods such as reversed-phase liquid chromatography coupled with tandem mass spectrometry (LC-MS/MS) (Yates, 2011). MS-based protein quantification can be achieved using label-based and label-free methods. Label-based quantitation methods make use of stable isotope labels that are incorporated within peptides to create an expected mass difference, whereas label-free methods use signal intensity and spectral counting of peptides for relative and absolute quantitation (Anand et al., 2017). Label-based methods include isobaric mass tags for relative and absolute quantitation (iTRAQ) (Wiese et al., 2007), and stable isotope labeling by amino acids in cell culture (SILAC) (Ong and Mann, 2007). Label-free methods include spectral counting, which are combined with downstream MS (Anand et al., 2017).

### *5.2. Structure-Function Relationships*

Beyond protein characterization and quantification, MS-based techniques can be used for structural and biophysical analysis at proteome scale, providing a holistic view of proteome-wide structural alterations, folding and stability, aggregation, and molecular interactions in cell lysates or intact cells (de Souza and Picotti, 2020). Using a massively parallel methodology combining size exclusion chromatography (SEC) to fractionate native protein complexes, SWATH/DIA mass spectrometry to precisely quantify the proteins in each SEC fraction, and a computational framework *CCProfiler*, the HEK293 proteome was interrogated to identify and quantify every protein complex in each cell (Heusel et al., 2019). Of the 4916 proteins identified and quantified, 2668 proteins (66%) and 55% of the protein mass was in an assembled state, organized as part of a macromolecular complex. While such complex-centric analysis has not been contemplated for production cell line development, it affords an unprecedented view of the structure-function relationships within a HEK293 cell that could inform strategies to improve productivity.

### *5.3. Enhancing Host Cell Productivity of rAAV*

Understanding the host cell proteome is necessary for devising engineering strategies that could deliver improvements in the host cell productivity. LC-MS/MS and relative label-free methods have been used to compare transfected versus non-transfected cells for AAV5-producing suspension HEK293 cells (Strasser et al., 2021). Proteins involved in cell organization, cell proliferation, and biogenesis were found to be the most upregulated in AAV5-producing cells, while proteins participating in metabolic processes were significantly downregulated. Strong upregulation was also reported in the Mannosidase Alpha Class 2A Member 2 (MAN2A2) protein in the cell pellet samples of the transfected HEK293 cells. MAN2A2 is known to be involved in N-glycosylation of proteins, which can critically influence the function and quality of recombinant proteins produced (Kaur, 2021; Rose, 2012). In addition to glycosylation being a critical quality attribute (CQA) of biotherapeutics, expression changes in certain mannosidases could be potential candidates for targeted modulation of certain PTMs in rAAV products. Functional enrichment and pathway analysis revealed differential expression patterns in endocytosis-lysosomal proteins that led to the hypothesis that blocking endocytosis might improve rAAV production. A follow-up experiment treating the cells with chloroquine, which is known to inhibit endocytosis, increased the viral titer by more than 35%. This work demonstrated proteomics utility combined with other systems biology approaches to gain understanding of the cellular processes and thereafter devise strategies for improving rAAV production.

### *5.4. Enhancing Host Cell Productivity of Virus-Like Particles*

ELISA, western blotting, and multiplexed quantitative proteomics via LC-MS/MS have been used to characterize protein profiles and cellular changes in HEK293 cells producing virus-like particles (VLPs) (Lavado-García et al., 2020b). Expression changes in cells grown under three operating conditions were investigated: (1) without transfection, (2) transient transfection of an empty plasmid, and (3) standard transient transfection of plasmid with Gag::EGFP gene. Transfection efficiency was observed to dramatically decrease at cell densities higher than ～$3\times10^6$ cells/mL, partially explained by a downregulation of intracellular protein transport to the nucleus and lipid biosynthesis. Further, the decrease in cellular viability upon transfection was caused by overall disruption of homeostasis due to multiple levels of homeostatic control being altered, such as calcium regulation, oxidant detoxification, DNA detoxification, glycosphingolipids metabolism, xenobiotic metabolism, and peptidase activity. Finally, during VLP production, while above traits were maintained, specific modifications in the membrane calcium channels and extracellular matrix were observed. Such proteomics-enabled findings could help devise future strategies to reduce bottlenecks and design less stressful processes, thereby increasing cellular productivity.

Follow-up work used LC-MS/MS to characterize extracellular vesicles (EVs) co-produced with VLPs (Lavado-García et al., 2020a). EVs were detected in all the three operating conditions as described above; however, during transfection, a shift in EV biogenesis was observed leading to changes in the size distribution of the EV products from large to small. Further functional analysis revealed that production of EVs reflected an overall energy homeostasis disruption via mitochondrial protein deregulation. Understanding the nature of the extracellular environment in VLP production will facilitate design of future VLP-based therapies, as it could lead to new downstream separation strategies or using copurified EVs as delivery vehicles.

These recent studies demonstrate the value of proteomic tools for studying the complexity of protein expression, function, PTMs discovery, and regulation in HEK293 cells.

## 6. Metabolomics

Biological processes involving complex interactions between genes, mRNA, proteins, and metabolites, combined with external environmental factors, have a profound influence on the resulting phenotype of the cell or organism. While genes determine what may happen, metabolites define what has happened, and are typically the final output of gene expression (Tan et al., 2016). Metabolites are small molecules serving as intermediates or end-products in cellular metabolic processes (Lamichhane et al., 2018). Metabolic flux is the passage of a metabolite through a pathway over time. A measurable change in a metabolic flux can occur in seconds to minutes, whereas genes and proteins take minutes to hours for measurable changes. Given that changes in the metabolome can occur rapidly, metabolites accurately reflect the state of disease or health at a given time point, and have been used as biomarkers to diagnose complex metabolic diseases or as novel therapeutic targets, and are hence valuable for precision medicine (Clish, 2015; Tan et al., 2016).

### *6.1. Metabolomic Analysis Techniques*

Metabolomics is broadly defined as the comprehensive measurement of all metabolites in a biological system, both endogenous and exogenous, that is, produced either internally by the system or derived from various external sources such as diet, microbes, or xenobiotic sources (Lamichhane et al., 2018). This approach is complementary to genomics, transcriptomics, and proteomics, and is the most recently developed omics technique that provides a sensitive measure of the phenotype of a biological system (Tan et al., 2016). High-throughput analytical techniques such as nuclear magnetic resonance (NMR) spectroscopy, gas chromatography (GC), and liquid chromatography coupled with mass spectrometry (LC-MS) are used in metabolomics to routinely measure tens to hundreds of metabolites with high accuracy and precision in a specimen, either through a targeted or non-targeted approach (Lamichhane et al., 2018). Fluxomics, on the other hand, describes the various approaches that seek to determine the complete set of metabolic fluxes in a cell or organism modeled through the metabolic network. Fluxomics can be considered a sub-branch of metabolomics that best represents the dynamic picture of the phenotype, resulting from the interactions of the metabolome genome, transcriptome, proteome, and epigenome (Cortassa et al., 2015; Emwas et al., 2022). Common fluxomics techniques are metabolic flux analysis (MFA), flux balance analysis (FBA), and $^{13}C$-fluxomics (Emwas et al., 2022). Fluxomics can be used to understand the current phenotype state and indicative of its next state(s) to help in development of control strategies for the host cells to produce highest titer and quality.

*6.2. Measuring the Metabolome*

In practice, metabolomics presents an analytical challenge because – unlike genomics, transcriptomics, and proteomics – metabolomics aims to measure molecules that have disparate physical and chemical properties. This requires the metabolites to be analyzed as distinct subsets, often grouped based on compound polarity, common functional groups, or structural similarity, using distinct sample procedures for each (Clish, 2015). In addition to the physical properties of the molecules and appropriate sample separation procedures, media composition has a significant influence on the outcome of metabolomic experiments. An examination of the influence of five different media combinations upon the cell secretome metabolite profiles of two phenotypically different cell lines – HEK293 and L6 rat muscle cells – found that the addition of fetal bovine serum (FBS) resulted in the detection of unique metabolites (Daskalaki et al., 2018). Media differences also had an impact on expression profiles, as glutamine and pyroglutamate were found to be more abundant in incubated relative to refrigerated medium. Among all the variables that can affect the secreted metabolite profile analyzed in the study, the choice of media was found to be the most sensitive.

*6.3. Metabolic Models*

Beyond the measurement of metabolites, a recent computational approach to map the entire metabolic information of an organism involved construction and use of genome-scale metabolic models (GEMs) (Bernstein et al., 2021). Similar to the role of a reference genome assembly in the study of genomics, transcriptomics, and proteomics, genome-scale metabolic models provide comprehensive, systems-level representations of cellular metabolic functions. GEMs depict complex cellular networks, integrating genes, enzymes, reactions, and metabolites. These models computationally map gene-protein-reaction associations, enabling flux predictions for diverse systems-level metabolic analyses (Bernstein et al., 2021; Gu et al., 2019; Passi et al., 2021). By providing a framework linking the organism's genome and gene expression to the metabolic phenotype, GEMs can be invoked to simulate cellular changes or perturbations and obtain a readout of the equivalent metabolic responses, as a unique fingerprint for the state of bioprocesses.

The first genome-scale metabolic model for human metabolism, RECON 1, was released in 2007, helping to unlock the complexity of the relationship between metabolites and cellular behavior. At the time, RECON 1 accounted for the functions of 1,496 open reading frames (ORFs), 2,004 proteins, 1,509 unique metabolites, and 3,744 metabolic, transport, and exchange reactions (Duarte et al., 2007; Thiele et al., 2013). After successive improvements and integration with Edinburgh Human Metabolic Network (EHMN), HepatoNet1, and other databases (Ryu et al., 2015), the initial version progressed into RECON 2 (Thiele et al., 2013), which included ~2x more reactions and ~1.7x more unique metabolites. RECON 3D, the most comprehensive human metabolic model, incorporates 3D metabolite and protein structure data. It encompasses 3,288 ORFs (17% of annotated human genes), 13,543 metabolic reactions, 4,140 unique metabolites, and 12,890 protein structures (Brunk et al., 2018).

Complementary to the RECON series, Human Metabolic Reaction (HMR) (Mardinoglu et al., 2014, 2013) is a well-established generic GEM series for human metabolism, focused on subcellular localization and tissue-specific gene expression (Gu et al., 2019). The HMR series, more comprehensive in fatty acid and lipid metabolism than RECON, resulted from extensive manual curation. It has spawned several cell-type specific GEMs, including iAdipocytes1809 (Mardinoglu et al., 2013), iHepatocytes2322 (Mardinoglu et al., 2014), and iMyocyte2419 (Väremo et al., 2016).

Multiple iterations of a metabolic model for HEK cells were constructed over the past two decades through meticulous manual curation efforts (Henry et al., 2005; Nadeau et al., 2002, 2000a, 2000b). Tissue- and cell-type specific omics data were integrated with a generic human metabolic model to derive a high-quality, context-specific metabolic model (Ryu et al., 2015). The HEK cell-type specific GEM was derived from a generic human model RECON 2 by devising a heuristic to emphasize flux through core metabolic reactions, and removing redundant pathways through a systematic, multi-step reduction process (Quek et al., 2014). This reduced model comprising only 357 active reactions describes the "functionalization" of RECON 2 for conducting steady-state metabolic flux analysis, and enables its application in investigating the metabolism of HEK cell cultures (Abbate et al., 2020; Martínez-Monge et al., 2019). For example, flux analysis based on a reduced HEK model (comprising 354 reactions and 335 metabolites when reduced from RECON 2.2 version using the same reduction protocol) evaluated the metabolic shift in HEK293 cell cultures from glucose consumption and lactate generation to glucose and lactate co-consumption during exponential growth (Martínez-Monge et al., 2019). This shift resulted in improved, balanced metabolism, with reduced glucose and amino acid uptake rates and minimal impact on cell growth. This insight could lead to new genetic engineering strategies that avoid the lactate accumulation in cell cultures. An alternative reduced HEK metabolic model with 82 reactions and 51 internal metabolites has been used to evaluate the comparative performance of adaptive flux variability analysis (AFVA) and classical flux variability analysis (FVA) in terms of coping with uncertainties arising from measurement noise and data smoothing (Abbate et al., 2020, 2019). The reduced HEK metabolic network and the biomass synthesis reactions were largely inspired from a metabolic network designed for CHO cells and not HEK cells (Fernandes et al., 2016). The biomass fluxes estimated using AFVA were reported to be more consistent than for FVA.

Metabolic flux analysis has also been helpful in discovering key metabolic pathways and their potential role in driving productivity improvements in HEK293. The underlying metabolic changes in pyruvate carboxylase 2 (PYC2)-expressing HEK293 cells compared to parental HEK293 cells have been investigated by estimating intracellular fluxes using various $^{13}C$-labeled substrates. Insertion of the PYC2 gene in HEK293 cells has led to marked reductions in lactate and ammonia production (which are considered toxic metabolic waste) during batch cultures, in addition to a 2-fold increase in maximum cell density and 33% increase in the yield of recombinant glycoprotein interferon-$\alpha$2b production, while maintaining maximum specific growth rate (Henry and Durocher, 2011). Flux analysis was performed to understand these improvements. The analysis indicated significant pyruvate carboxylase activity in the PYC-expressing cells, that was previously missing in the parental cells. Additionally, key flux differences in the PYC-expressing cells came to light, for example, at the pyruvate branch point, that the cells converted 56% of the pyruvate pool into acetyl-CoA (compared to 51% in parental cells), 26% into oxaloacetate via the pyruvate carboxylase (not reported in parental cells), and only 9% into lactate (compared to 48% via lactate dehydrogenase in the parental cells). Flux analysis also indicated an overall reduction in glucose uptake rate in PYC-expressing cells, and found that the amount of glucose channeled through the pentose phosphate pathway (PPP) was reduced, representing 7% of the glucose uptake rate, compared to 28% for the parental cells. Such detailed insights can help guide future cellular engineering approaches and design of improved feeding strategies and/or media formulations to maximize host cell productivity and enhanced product quality.

### *6.4. Metabolomics for Optimal Culturing Conditions*

Metabolic flux analysis has also been used to determine HEK293 perfusion culture conditions that enhanced productivity of adenoviral vectors (Henry et al., 2005). Intracellular flux analysis showed that experiments with highest specific product yields displayed increased glycolytic and TCA cycle fluxes, along with enhanced ATP production rates at infection time. A feeding strategy was established that maintained favorable metabolic state throughput during the growth phase and allowed successful infection at a cell density of $5{\times}10^6$ cells/ml, resulting in specific productivity close to the maximum achieved while having a 2-fold increase in cell density.

### *6.5. Lipidomics*

Lipidomics, a subfield of metabolomics focused on the lipidome, is relatively understudied and underutilized. Lipids are potentially important regulators of protein production and secretion because of their involvement in energy metabolism, vesicular transport, membrane structure, dynamics, and signaling (van Meer et al., 2008). Combined lipidomic and transcriptomic analysis of HEK293, CHO, and mouse myeloma SP2/0 cell lines revealed that phosphatidylethanolamine (PE) and phosphatidylcholine (PC) are the major lipid components in all three systems (Zhang et al., 2017). HEK293 cells had 4–10 times more lysophosphatidylethanolamine (LPE) and 2–4 times more lysophosphatidylcholine (LPC) than CHO and SP2/0 cells, consistent with the transcription levels of the relevant enzymes. As previously mentioned, indirect lipidomic analysis using transcriptomic and metabolomic gene analysis to pinpoint differences between adherent and suspension HEK293 cell lines, revealed pathway changes related to cholesterol biosynthesis and metabolism (Malm et al., 2020). Transcriptomic analysis of three HEK293 cell lines (high, intermediate, and low producers) showed that cholesterol biosynthesis levels were inversely correlated with rAAV production efficiency, suggesting that high cellular cholesterol levels may be disadvantageous (Pistek et al., 2023). Given that cholesterol is an important component of the cell membrane and is integral to membrane structure, cell signaling, and protein trafficking, its impact on plasmid transfection and rAAV synthesis warrant investigation. In the future, systematic lipidomic analysis could help elucidate the physiological and secretory machinery of HEK293 involved in rAAV production to inform improvements in process design and cell line engineering.

Metabolomics and metabolic models provide actionable insight for engineering cells to achieve optimal cell membranes for secretion of product and minimize sensitivities to suspended cell processing (e.g., shear from mixing, gassing, aggregation).

## 7. Multi-omics

Applications of omics such as genomics, transcriptomics, proteomics, and metabolomics have, independently, revealed valuable insights into the cellular mechanisms of mammalian host cell systems such as CHO and HEK293 for producing recombinant proteins. Single omics approaches, although effective, are overly simplistic and fall short in explaining the complex intracellular and systems biology workings of a cell or organism. Combining different omics data in the form of multi-omics analysis can bring together different biological layers and components in a cell to synthesize a more complete mechanistic understanding. Using multi-omics, insights drawn from genomics can be connected to gene expression differences through transcriptomics, for example, and, in other cases, help understand complex molecular pathways that would be difficult or impossible to ascertain with single-omics approaches. In this section, we highlight examples of multi-omics analysis applied to HEK293 cell systems and the molecular insights that they reveal for recombinant protein production, while highlighting the challenges and opportunities that lie ahead.

### *7.1. Multi-omics for Cell Line Characterization*

Combining genomics, transcriptomics, and metabolic pathway analysis, multi-omics was used for cell line characterization of 6 widely used HEK293 cell lines. In addition to providing a cross-validation of results, this holistic approach revealed fundamental evolutionary differences for adapting to industrial production cell lines (Malm et al., 2020; see Sections 2, 4, and 6). Notably, key common changes between HEK293 and its progeny production cell lines involved integral membrane proteins and processes related to cell adhesion, motility and the organization of various cellular components such as the cytoskeleton and extracellular matrix.

In addition to shedding light on key morphological and biochemical changes, multi-omics was used to detect and confirm unanticipated (and potentially undesirable) changes. For example, the detection of unexpected expression of the large T antigen gene in a HEK293E cell line from the transcriptomics data was confirmed using complementary genomics data, which revealed the presence of a partial large T antigen gene in the 293E genome that was previously unreported. Tracing the origin of the HEK293E cell line postulated that this large T antigen may have come from plasmids used to co-transfect the HEK293 cell line.

Understanding these evolutionary differences between parental HEK293 and its derivatives can guide enhanced suspension cell line engineering and process design for improved rAAV production for clinical and commercial supply.

### *7.2. Multi-omic Analysis of Cellular Dynamics and Responses*

By integrating transcriptomic with targeted quantitative proteomic analysis, Lu and colleagues provided new insights into the biology and kinetics of rAAV production in HEK293 (Lu et al., 2024). Using RNA-Seq, absolute quantification through targeted proteomics (AQUA), and tandem mass tags (TMTs), analysis of 20,000 unique transcripts and 5000 proteins was used to measure the kinetics of rAAV triple transient transfection over time (0, 24, 48, and 72 hours post-transfection). This analysis revealed that only 5.7% total genomes (TG) per cell became encapsidated as full viral genomes (VG), suggesting that capsid assembly, genome encapsidation, or both are likely limiting factors for rAAV production. This is consistent with the findings of Braatz and colleagues, whose mechanistic model for rAAV production attributed the low filled capsid fraction to a poorly coordinated timeline between capsid synthesis and viral DNA replication, and the repression of later phase capsid formation by Rep proteins (Braatz et al., 2024).

The same study leveraged transcriptomics and proteomics to functionally assess cellular responses, shedding new insights into known responses to rAAV production in HEK293 that negatively or positively impact productivity (e.g., inflammation and the host innate immune response, the antiviral response, the unfolded protein response, MAPK cascade, cell cycle inhibition, DNA damage). Further, it revealed that certain host defense responses were a general consequence of plasmid DNA transfection that was independent of rAAV production. By providing a nuanced systematic mechanistic understanding of these complex phenomena, multi-omics enables a precise means to increase productivity through chemical intervention (e.g., by the addition of specific inhibitor or enhancer molecules) or through synthetic biology (e.g., by engineering the HEK293 host cell line or the rAAV vector).

### *7.3. Insights into Production Constraints*

Multi-omics analysis of HEK293 stable producer cells revealed broad cellular network adaptations that likely support recombinant protein production (Dietmair et al., 2012). Transcriptomics, metabolomics, and fluxomics data showed downregulation of genes involved in cell growth and proliferation, explaining how producer cells maintained growth rates despite lower glucose consumption and the burden of recombinant protein production. Simultaneously, upregulation of genes related to endoplasmic reticulum stress suggested constraints in protein folding and assembly.

Through a comparative transcriptomics and proteomic analysis, bottlenecks in rAAV production were investigated to enhance the genetics and processes for producing high-quality viral vectors. Using RNA-seq coupled with AQUAs, a HEK293 stable producer cell line (PCL) carrying EGFP was compared with two wtAAV production systems, one by multi-plasmid transient transfection and the other by virus co-infection (Lin et al., 2024). Genes involved in inflammation, innate immunity, antiviral response, and MAPK signaling were upregulated in all three scenarios, while those involved in mitochondrial function were downregulated, identifying potential targets for genetic intervention in both transiently stably transfected HEK293 cell lines.

While all three systems produced comparable copies of AAV genomes, the kinetics of viral capsid production differed dramatically, resulting in a 10-fold lower titer for the PCL due to capsid insufficiency and inefficient genome packaging. Another notable difference between the stable PCL versus the two transient systems was that the GFP cargo in the PCL was the most abundant transcript of all cellular mRNA, despite being under the control of an inducible LacSwitch. Constituting ~20% of total cellular transcripts at 72 hours, this undesirable cargo expression is a significant waste of cellular resources within the stable cell line. While the scalability and low-cost potential of rAAV PCLs are appealing, they still significantly underperform transient transfection in terms of titer and yield, despite major investment in time and resources over 30+ years. Furthermore, cell line stability is unproven (Escandell et al., 2022; Shupe et al., 2022), and not a single rAAV product produced in a PCL has entered the clinic to date, reflecting caution across the industry.

### *7.4. Integrating Three or More Omics*

While combining two branches of omics is powerful for uncovering actionable information for improving production in HEK293, integrating data from three or more omic disciplines can reveal additional useful insights. This has proven to be a successful strategy for understanding complex biological processes, including constraints of recombinant protein production in HEK293. To our knowledge, there are no examples of an integrated multiomics approach for producing rAAV from HEK293 cells, therefore, we look to examples producing lentivirus and other recombinant proteins. For example, a functional genomics study used transcriptional profiling and central carbon metabolism analysis to identify pathways and bottlenecks in the production of lentivirus by investigating the transition from parental to producing cell lines (Rodrigues et al., 2013). The cell lines investigated were two human-derived cell lines, HEK293 FLEX (low producer) and Te Fly (high producer), and their parental cell lines (HEK293 and Te671, respectively). Since the producing cell lines have different transcriptomes, it was difficult to find a common expression signature to compare the producing vs. parental cell lines through a single omics approach. Therefore, applying three omics approaches to evaluate genes, gene expression changes, and metabolites across the four cell lines helped identify eight distinct pathways to be recruited in the virus production state. To confirm their findings, the culture medium was supplemented to positively influence these pathways and results showed improved titer by up to 6-fold. Further, by investigating two producing cell lines, with different genetic backgrounds, several genes were identified as possible targets for future engineering strategies as they were limiting production in the low production phenotype. Applying and integrating multiple omics approaches proved to reveal actionable insights into the pathways involved in lentiviral production, such as demonstrating that manipulation of nucleotides and polyamine metabolism through media supplementation can be used to increase productivity.

In a separate study, a stable HEK293 producer cell line was compared to its non-producing parental cell line using transcriptomics, metabolomics, and fluxomics, including flux variability analysis (FVA) to measure the metabolite consumption and production rates (Dietmair et al., 2012). Integrating multiple omics approaches revealed the producer cells consumed less glucose than the non-producing cells and the majority of genes involved with oxidative phosphorylation were downregulated, with no consistent evidence of glycolytic energy compensation. Dietmair and colleagues speculated that a broad adaptation of the cellular network freed resources for other mechanisms to rise for production of recombinant proteins. Additionally, transcript abundance often correlated positively with fluxes and metabolite concentrations. However, some discrepancies were also noted between transcript levels and fluxes, with occasional counterintuitive outcomes like increased glucose transporter expression paired with reduced glycolytic flux in producer cells. Comparing cellular behavior at multiple levels enabled by an integrated omics approach revealed techniques for rational engineering of cell lines to improve production.

### *7.5. Multi-omics for Mechanistic Understanding*

Multi-omics enables a unique comprehensive view of the underlying mechanisms employed by the host cell system, building on decades of biochemistry, genetic, and cell biology that informs our current mechanistic understanding of recombinant protein production in HEK293. Highlights into the mechanisms employed by the host cell system have been described throughout this review. For example, as previously mentioned, Tworig and colleagues conducted a transcriptomics analysis resulting in mechanistic knowledge of druggable pathways in HEK293 cells for an informed experiment using small molecules to regulate core cellular functions (Tworig et al., 2024) (see Section 4).

Additionally, transcriptomics has been combined with high-throughput lipidomics to compare HEK293 with two other widely used mammalian expression systems, CHO and SP2/0, during both exponential and stationary growth phases (Zhang et al., 2017) (see Section 6). Analyzing transcriptomic expression profiles together with lipidomics data led to four mechanistic findings: (1) LPE and LPC levels were significantly higher in HEK293 cell lines, likely promoting membrane curvature, which can facilitate membrane tubule formation and membrane vesicle fusion and fission, (2) lysophospholipids (LPLs) may play an close role in exocytosis and secretion processes, (3) group IV sPLA2 enzymes are unique to the HEK293 cell line, which may enhance its intracellular membrane trafficking and fusion events, and (4) further promotion of membrane trafficking and tubule formation events may be contributed to low expression of lysophospholipid acyltransferase (LPEAT) and lysophosphatidylcholine acyltransferase (LPCAT) in HEK. This study demonstrates the power of integrating multiple omics approaches for elucidating the mechanisms that limit protein expression, exocytosis, and secretion in HEK293 cells.

While small, rationally designed multi-omics studies can lead to valuable mechanistic insights, mechanistic models provide the most robust actionable information for furthering rAAV production in HEK293 cells (Masson et al., 2023). Informed by omics data, mechanistic models can encompass a range of information from the genome scale to bioreactor scale, and from cell culturing to protein production (Canova et al., 2023; Masson et al., 2023). For those phenomena in which the existing mechanistic phenomena are not sufficiently well understood to build purely mechanistic models, data analytics/machine learning (DA/ML) tools can be applied to build those relationships, which can be integrated with the mechanistic model components to produce hybrid models (Tsopanoglou and Jiménez del Val, 2021). A general systematic modeling framework for building hybrid models from multi-omics data, however, is not yet available.

Mechanistic modeling has been an invaluable tool to support process development for biotherapeutic production in CHO cells, and more recently, HEK293 cells (Sha et al., 2018). For example, Braatz and colleagues developed a mechanistic model of rAAV via transient triple transfection of HEK293 cells growing in suspension has been developed and used to identify bottlenecks and propose improvements to the system (Nguyen et al., 2021). As previously mentioned, application of this model provided insight into a timeline disconnect between capsid synthesis and viral DNA replication, ultimately leading to a novel multi-dose transfection method to coordinate these timelines (Braatz et al., 2024). Additionally, utilizing the underlying mechanisms to produce AAV incorporated in the mechanistic model, a first-in-class mechanistic model for continuous AAV production via transfection in a perfusion bioreactor was constructed (Nguyen et al., unpublished results). The model was used to design a continuous process with high cell density transfection and experimentally validated the approach of using multiple dosing for re-transfection. The resulting high-density transfection and re-transfection resulted in increased batch runtime efficiency by 8-fold (i.e., 8 times the amount of a conventional batch in the same amount of time) and increased titer production per unit plasmid by ~32%. Integrating mechanistic models with omics datasets has untapped potential in identifying bottlenecks that may not be obvious from the omics datasets alone; therefore, furthering the toolbox of mechanistic understandings for production of rAAV from HEK293 cells (Lazarou et al., 2020; Masson et al., 2023).

### *7.6. Challenges and Opportunities*

Analysis of multi-omics data can be profoundly insightful in understanding the cellular mechanisms; however, the extraction of meaningful biological knowledge from complex large datasets representing different layers of the cellular machinery can be challenging. The process of generating biological knowledge typically involves four key steps: data generation, processing, integration, and analysis (Palsson and Zengler, 2010). Although today's low costs of sequencing and automated bioinformatics pipelines make the first two steps fairly straightforward for most omics data, the third and fourth steps can be difficult in delivering data-driven insights as multi-omics data differ in type, scale, and distribution with thousands of variables and only few samples (Picard et al., 2021). For preliminary insights, cross-omics data integration can be achieved through basic statistical approaches such as correlation analysis across enriched biological features from individual omics datasets. However, to extract deeper and more profound and nuanced insights, artificial intelligence (AI)-based approaches, in particular machine learning (ML), are increasingly applied (Kang et al., 2022; Reel et al., 2021). By being able to better model the inter-omics interactions between complementary omics datasets, ML approaches have the potential to increase understanding of the underlying biological mechanisms. This has been demonstrated by the recent surge in their application for large-scale multi-omics analysis of cancer datasets (Chaudhary et al., 2018; Li et al., 2022; Nicora et al., 2020; Opdam et al., 2017; Poirion et al., 2021; Wang et al., 2021). In the context of biotherapeutic production, ML has been applied for genome-scale modeling of CHO metabolic networks (Schinn et al., 2021; Zampieri et al., 2019) and for predicting protein glycosylation patterns in CHO cells (Antonakoudis et al., 2021; Kotidis and Kontoravdi, 2020).

As new tools emerge for improving rAAV and recombinant protein production, multi-omics can be leveraged to elucidate the underlying mechanism and further improve productivity. The recent emergence of small-molecule (SM) chemical additives is a case in point, where a growing number of SMs has been empirically shown to increase rAAV production (Maznyi, 2023; Reese et al., 2023.; Scarrott et al., 2023). One such class of SMs, known as viral sensitizers, enhance viral production by transiently attenuating antiviral pathways in the cell, such as the innate immune response (Diallo and De Jong, 2023). With a growing number of commercially available SM additives for enhancing viral vector manufacturing, multi-omics analysis could enable the systematic elucidation of the mechanism of action for each molecule, and shed light on the pathways and kinetics. Such omics-enabled insights could shed light on potential synergies, inform new mechanistic models for rAAV production, and fuel new engineering strategies to improve HEK293 cell lines, such as attenuating the innate immune response.

The ever-growing ability to generate a panoply of large data sets from multiplicity of sources and techniques combined with the unprecedented power of AI-enabled approaches can lead to the temptation to apply AI/ML to analyze the immediate data to look for patterns. However, this simplistic tendency overlooks decades of information about fundamental biological processes and leads to minor incremental improvements, at best. To gain novel, useful, and actionable insights, multi-omics should be combined with additional layers from our “institutional knowledge.” Today, the analysis and integration of multi-omics data is still relatively untapped for HEK293, and therein lies an opportunity. Doing so could enable exponential improvements in productivity that would have a major impact on rAAV gene therapy production in terms of yield, speed, and cost (Fig. 1).

## 8. Conclusion and Future Directions

For rAAV gene therapy to cross the chasm from niche to mainstream medicine, there must be a step change in scalable, cost-effective production. We propose exploiting HEKomics – ranging from single- to multi-omics – to address knowledge gaps and identify opportunities to optimize rAAV production by revealing actionable insights (Fig. 1 and Table 1). This section discusses some opportunities beyond omics and HEK293 for scalable production of rAAV-based gene therapies.

Relative to genomics, transcriptomics, proteomics, epigenomics and metabolomics, the area of glycomics in HEK293 is the least explored. Given the impact of N-linked glycosylation on the potency, safety, immunogenicity, and pharmacokinetic clearance of therapeutic proteins including monoclonal antibodies (Heffner et al., 2021; Narimatsu et al., 2019), this is a critical knowledge gap. As a case in point, integrating glycome data with transcriptomic and metabolomic data measured during a time course of mAb production in a CHO fed-batch culture revealed that the steps involving galactose and sialic acid addition were temporal bottlenecks (Sumit et al., 2019).

The relevance of glycomics to rAAV production in HEK293 was underscored by the glycomic profiling of capsid proteins from AAV serotypes (Xie and Butler, 2023). Using a high-sensitivity N-glycan profiling platform, analysis of 9 serotypes (AAV1-9) showed that all had comparable glycosylation profiles, characterized by high mannosylated N-glycans and lower fucosylated and sialylated N-glycan structures. However, the precise glycan compositions differed across the different serotypes. Given that many virus-host interactions are driven or influenced by glycans (Miller et al., 2021; Zhou et al., 2024), these data suggest that the differences could play a role in tissue tropism, cell surface receptor binding, intracellular uptake and processing.

Leveraging transcriptomics, Huang and colleagues developed GlycoMaple, a glycosylation mapping tool for visualizing glycosylation pathways and estimating glycan structures in human-derived cells (Huang et al., 2021). This tool was verified using HEK293 cells, determining 38 and 14 composition structures in N-glycan and O-glycan analysis, respectively. Characterization of these glycan structures further informed engineering strategies to knock out certain genes involved in LacdiNAc formation to eliminate LacdiNAc structures. This tool has proven to be useful in predicting changes in glycans to genetic mutations, but further experimentation can be completed to investigate limiting or critical steps in rAAV glycosylation pathways.

Beyond glycomics, epigenomics and the role of epigenetics in HEK293-producing cell lines are relatively unexplored to date. Methylation and chromatin analysis of CHO cell lines (Hilliard and Lee, 2021; Moritz et al., 2016; Osterlehner et al., 2011; Patel et al., 2018), combined with preliminary analysis of rAAV production in HEK293, points to epigenetic engineering as a promising means to improve productivity. Further, epigenetics can potentially be driving the adaptation of HEK293 cells to serum-free suspension medium from adherent medium, as in the case of CHO cells (Feichtinger et al., 2016), and yet a comprehensive epigenomics study investigating these mechanisms in AAV-producing HEK293 cells is missing.

When exploring approaches to solve the fundamental gap in scalable rAAV production, it is prudent to consider alternatives. The Sf9 insect cell baculovirus expression vector system (BEVS) has emerged as a viable alternative rAAV production system (Joshi et al., 2021), and the first BEV-based gene therapy (HEMGENIX®) was approved in 2022 (Herzog et al., 2023). While the BEVS system has several advantages over HEK293 (Destro et al., 2023), many challenges remain that would also benefit from insights through omics analysis. Transcriptome analysis of dual-baculovirus infection of Sf9 cells revealed genes that were differentially expressed during rAAV production, revealing potential engineering targets for process engineering (Virgolini et al., 2023a). Also, mechanistic models have been developed to describe the production dynamics of rAAV in the One-Bac, Two-Bac, and Three-Bac BEVSs and applied to identify bottlenecks that limit full capsid formation (Destro et al., 2023). Integrating the mechanistic models for rAAV production in Sf9 cells with the corresponding omics data could improve their predictive accuracy and broaden their utility.

Since little is known about the variability of insect cell populations and the potential effects of heterogeneity on viral titer and/or quality, single-cell RNA sequencing (scRNA-seq) was performed to shed light on transcriptional variation within a cell population following infection (Virgolini et al., 2023b). At 24 h post-infection, only 29.4% of cells expressed all of the three rAAV transgenes needed to produce full rAAV capsids. This study highlights the power of single-cell omics, revealing production bottlenecks and how to address them, such as using synchronization strategies to reduce cell heterogeneity. It also underscores an intrinsic limitation of current omic-based analysis: most approaches measure averages across cell populations and limited number of time points, obscuring the stochasticity and temporal variation for every step of a process, particularly one as complex as rAAV production. As single-cell multi-omics technologies improve in terms of throughput, data analytics, and affordable pricing (Baysoy et al., 2023; Heumos et al., 2023), they will become increasingly used to characterize cell-to-cell variability and use that information to increase overall system productivity.

Finally, in light of the HEK293 productivity gap for rAAV gene therapy production, it bears re-examining why CHO was not adopted as the system of choice. There is speculation that cellular restriction factors in CHO cells could affect rAAV production and interfere with virus packaging, as reported for other viruses (Nagy et al., 2023). Nevertheless, two new CHO-based systems for rAAV production recently came to light. Nagy and colleagues engineered a stable CHO cell line to express entry receptors so that they can be transfected using the herpes simplex virus (HSV) (Nagy et al., 2023). Although the titers resulting from transfection of the CHO cells were low (1.6–$4.1\times10^9$ vg/mL obtained 24 hours post-transfection), the resulting rAAVs had comparable *in vitro* transduction, infectivity, and biodistribution titers to rAAVs produced by triple transient transfection of HEK293 cells. In contrast to the stable cell line approach, Cao and colleagues developed a helper-free, transient 5-plasmid transfection CHO system for rAAV production (Cao et al., 2023). Despite low productivity, the study identified cap protein expression as a limiting factor for rAAV production in CHO cells, laying the groundwork for future research. While one of the themes of this article is that HEK-omics can learn from studies on CHO-omics, the contrast is true for increasing titer for rAAV production in CHO. To borrow from Shakespeare, the wheel has come full circle.

## Declaration of competing interest

The authors declare that they have no known competing financial interests or personal relationships that could have appeared to influence the work reported in this paper.

## Acknowledgements

This work was supported by an EPoC Grant from the Delaware Bioscience Center for Advanced Technologies to SWP in collaboration with ITR. GG received additional support from a Doctoral Fellowship for Excellence from the University of Delaware Graduate College.

## Tables

**Table 1.** HEK-Omics: Gaps and Opportunities

| Omics Type | Gaps and Opportunities | Key References |
|---|---|---|
| Genomics | • Establish comprehensive HEK293 cell line-specific reference map and gene annotation set.<br>• Investigate genetic changes in the form of SNPs, CNVs, and structural variants between HEK293 clonal populations with different phenotypes, e.g., linking varying titer productivity levels to genetic characteristics. | (Lin et al. 2014; Malm et al. 2020) |
| Epigenomics | • Characterize epigenetic differences among HEK293 cell lines under various culture and media conditions, e.g., cell cultures adapted to suspension versus adherent medium.<br>• Investigate unpredictable outcomes occurring in HEK293 cells producing rAAV, e.g., cell density effect. | (Broche et al. 2021; Tóth et al. 2019; Chanda et al. 2017; Feichtinger et al. 2016; Marx et al. 2022) |
| Transcriptomics | • Investigate transcriptional changes in transiently transfected HEK293 cell cultures and identify key regulatory pathways invoked during the production of recombinant AAVs.<br>• Investigate druggable pathways within HEK293 cells to enhance transfection and optimize small-molecule usage. | (Chung et al. 2023; Malm et al. 2020, 2022; Saghaleyni et al. 2022; Wang et al. 2023; Tworig et al. 2024) |
| Proteomics | • Identify and utilize structure-function relationships for enhancing engineered HEK293 producing cell lines.<br>• Analyze proteomic characteristics and differences among widely used AAV-producing HEK293 cell lines, e.g., HEK293E, HEK293T (adherent lines), HEK293H, HEK293F, and Freestyle HEK293F (suspension lines).<br>• Proteomic profiling of AAV-producing HEK293 cell lines in manufacturing environments to study the effects of process conditions on productivity. | (Strasser et al. 2021; Lavado-García, Jorge, et al. 2020; Malm et al. 2020; Huesel et al. 2019) |
| Metabolomics | • Develop comprehensive metabolic model for HEK293 cells, derived from human genome-scale metabolic models such as RECON or HMR series. | (Henry et al. 2005; Henry and Durocher 2011; Martínez-Monge et al. 2019; Quek et al. 2014) |
| Multi-omics | • Multi-omics study of HEK293 cells during rAAV production process to identify host cellular mechanisms favorable for higher productivity and better quality (to parallel similar studies for CHO, e.g., Lee et al., 2021).<br>• Integrated multi-omics study of HEK293 cells producing rAAV using 3 or more omics approaches.<br>• Develop mechanistic or hybrid models for HEK293 producing rAAV. | (Dietmair et al. 2012; Malm et al. 2020; Nguyen et al, 2021; Lu et al., 2024) |

## Figures

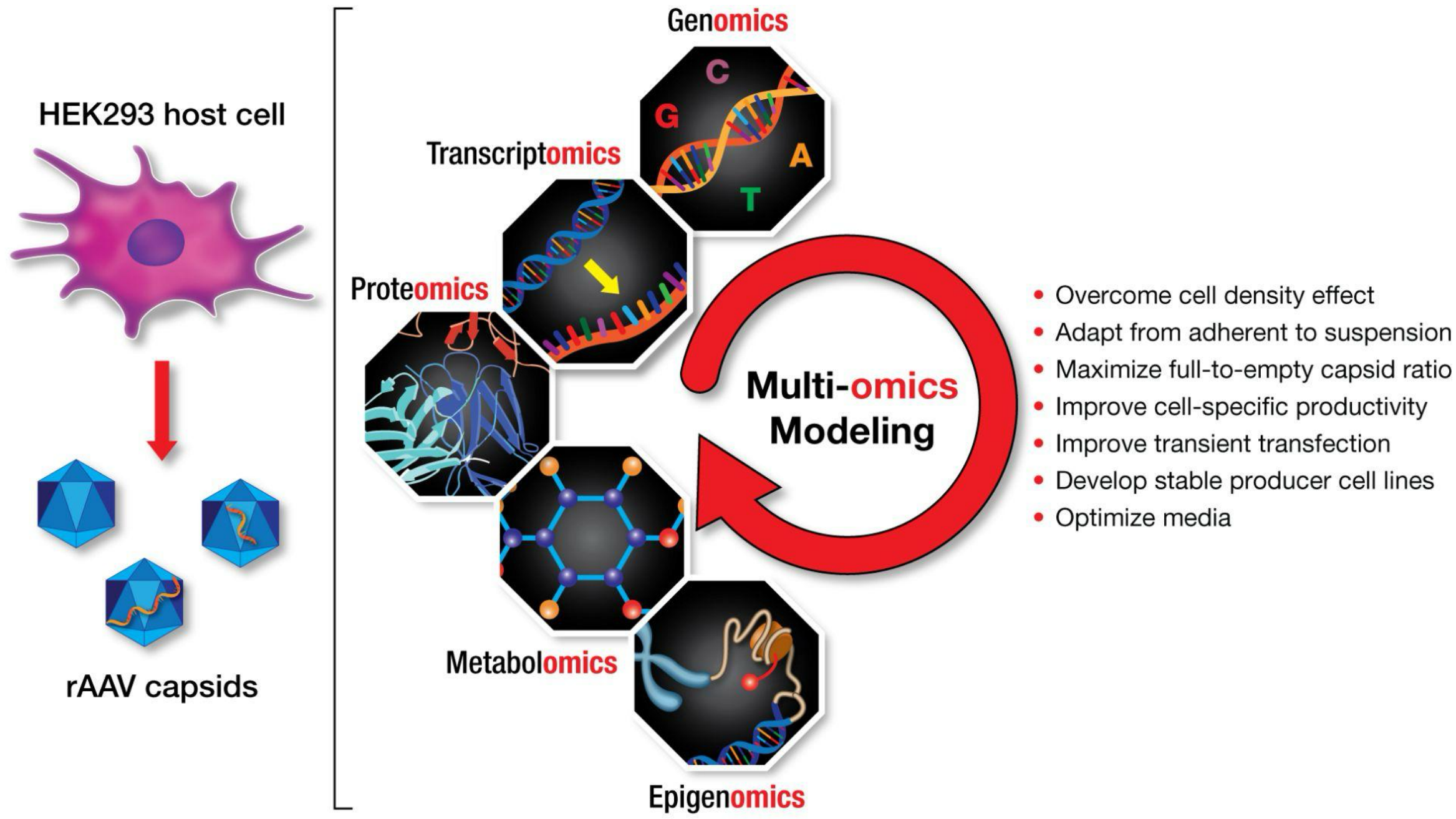

**Figure 1.** Potential opportunities to improve rAAV production in HEK293 cells through application of multi-omics technologies.